\documentclass[prd, twocolumn, nofootinbib, floatfix]{revtex4}

\usepackage{epsfig} 
\usepackage{amsmath}
\usepackage{amsbsy,color}

\newcommand{\beq}{\begin{equation}}
\newcommand{\eeq}{\end{equation}}
\newcommand{\beqa}{\begin{eqnarray}}
\newcommand{\eeqa}{\end{eqnarray}}
\newcommand{\om}{\Omega_m}

\newcommand{\ome}{\Omega_e}
\newcommand{\ode}{\Omega_{de}}
\newcommand{\rhoi}{\rho_\infty}

\newcommand{\gs}{\gtrsim} 
\newcommand{\ls}{\lesssim}

\def\ha{\frac{1}{2}}
\def\imag{\dot \imath}

\def \h{\mathcal{H}}
\def \sn2{\left(S/N\right)^2}

\begin{document} 

\title{Measuring the Speed of Dark: Detecting Dark Energy Perturbations}
\author{Roland de Putter$^{1}$, Dragan Huterer$^2$, Eric V.\ Linder$^{1,3}$
\vspace{0.1cm}
}

\affiliation{$^1$Berkeley Lab \& University of California, Berkeley, CA 94720, USA \\ 
$^2$Department of Physics, University of Michigan, 450 Church St, Ann Arbor,
MI, 48109, USA\\ 
$^3$Institute for the Early Universe, Ewha Womans University, Seoul, Korea} 
\date{\today}

\begin{abstract} 
The nature of dark energy can be probed not only through its equation 
of state, but also through its microphysics, characterized by the sound speed 
of perturbations to the dark energy density and pressure.  As the 
sound speed drops below the speed of light, dark energy inhomogeneities 
increase, affecting both CMB and matter power spectra.  We show that 
current data can put no significant constraints on the value of the 
sound speed when dark energy is purely a recent phenomenon, but can 
begin to show more interesting results for early dark energy models.  For example, the best fit model 
for current data has a slight preference for dynamics 
($w(a)\ne-1$), degrees of freedom distinct from quintessence ($c_s\ne1$), 
and early presence of dark energy ($\Omega_{\rm de}(a\ll1)\ne0$). 
Future data may open 
a new window on dark energy by measuring its spatial as well as time 
variation. 
\end{abstract} 

\maketitle

\section{Introduction \label{sec:intro}}

Although dark energy dominates the energy density of the universe and drives
the accelerating cosmic expansion, we know remarkably little about it. Over
the course of the past decade, cosmologists have devoted considerable effort
to devising new and sharpening known methods for determining the equation of
state of dark energy.  The equation of state, defined as the pressure to energy
density ratio, is generally a time dependent function and fully specifies 
the temporal evolution of dark energy density.  The dark energy density in 
turn (along with the matter density) determines the expansion rate of the 
universe, as well as geometrical measures (distances and volumes).

The equation of state $w(z)$ does not, however, tell us about the 
microphysics of dark energy, nor does it describe all of the cosmological 
signatures.  For example, even a perfectly measured $w(z)$ does not tell 
us whether dark energy arises from a canonical, minimally coupled scalar 
field, a more complicated fluid description, or modification of 
gravitational theory on large scales.  The properties of the perturbations 
to the dark energy, which must exist unless it is simply a cosmological 
constant or only an effective field, do carry such extra information. 

Perturbations to the energy density and pressure can be described through the
sound speed, $c_s^2=\delta p/\delta\rho$.  The sound speed carries information
about the internal degrees of freedom: for example, rolling scalar fields
(quintessence) necessarily have sound speed equal to the speed of light,
$c_s=1$.  Detection of a sound speed distinct from the speed of light would
indicate further degrees beyond a canonical, minimally coupled scalar field.

A low sound speed enhances the spatial variations of the dark energy, giving
inhomogeneities or clustering.  Heuristically, the sound speed determines the
sound horizon of the fluid, $l_s = c_s/H$, where $H$ is the Hubble scale.  On
scales below this sound horizon, the fluid is smooth; on scales above $l_s$,
the fluid can cluster. Since for quintessence $c_s=1$, the sound horizon
equals the cosmic horizon size and there are essentially no observable
inhomogeneities.  However, if the sound speed is smaller, then dark energy
perturbations may be detectable on correspondingly more observable (though
typically still large) scales.  These perturbations act in turn as a source
for the gravitational potential, and affect the propagation of photons. For
example, clustering dark energy influences the growth of density fluctuations
in the matter, and large scale structure, and an evolving gravitational
potential generates the Integrated Sachs-Wolfe (ISW) effect
\cite{Sachs:1967er} in the cosmic microwave background.  The observational
signatures of these effects offer a way of
probing the dark energy inhomogeneity and sound speed.

In this paper we study the signatures of the sound speed of dark energy. We
revisit and extend previous studies of dark energy clustering
\cite{Huetal98,Hu_synergy,Erickson01,Dedeo2003,Weller_Lewis,BeanDore04,Afsh04,HuScran04,Hannestad2005,Corasaniti2005,Koivisto_Mota_2005,Takada2006,Xia2007,Ichiki2007,Mota2007,Bashinsky2007,TorresRodriguez,Dent2008,Ballesteros_Riotto,SapKun09},
clarifying and quantifying the physical effects caused by the nonstandard 
values for the speed of sound. We then study models where the dark energy 
density was non-negligible at early times, which offer much better prospects 
for observable $c_s$ signatures than the fiducial near-$\Lambda$CDM case.  
Finally, using current cosmological data, we
constrain the speed of sound jointly with 7-8 other standard
cosmological parameters.

This paper is organized as follows.  In Sec.~\ref{sec:depert} we describe dark
energy perturbations and the physical influence of the sound speed and
equation of state, deriving the dark energy density power spectrum.  
Section~\ref{sec: models} describes the dark energy models we consider, and
Sec.~\ref{sec:cmbgal} treats the impact of dark energy inhomogeneity on the
CMB, matter power spectrum, and their crosscorrelation.  We consider models
with both constant and time varying equation of state and sound speed in
Sec.~\ref{sec:data}, and present constraints from current data.

\section{Dark Energy Perturbations \label{sec:depert}}

We briefly review the growth of density perturbations, in both the 
matter and dark energy, focusing on the role of the sound speed. 
See \cite{Kodama84, MaBert95} for more details. 
To derive the influence of the sound speed on dark energy inhomogeneity, 
and dark energy perturbations on the matter distribution, we must solve 
the perturbed Einstein 
equations for the density perturbations $\delta\rho_i$, pressure 
perturbations $\delta p_i$, and velocity (divergence) perturbations 
$\theta_i$.  We do not consider an anisotropic stress.  

In the conformal Newtonian gauge, the perturbed Friedmann-Robertson-Walker 
metric takes the form 
\begin{equation}
ds^2 = a(\tau)^2 \left[-(1+2\psi) d\tau^2 + (1-2\phi) d\vec{r}\,{}^2\right],
\end{equation} 
where $a$ is the scale factor, $\tau$ is the conformal time, $\vec{r}$
represents the three spatial coordinates, and $\psi$ and $\phi$ are the metric
potentials. Conservation of the stress-energy tensor 
($T^{\mu \nu}_{\,\,\,\,\,;\nu} = 0$) of a perfect fluid gives the following 
equations in Fourier space (see, e.g., \cite{MaBert95}) from the time-time 
and space-space parts: 
\begin{eqnarray}
\label{eq:continuity}
\frac{\dot{\delta}}{1 + w} &=& - \theta + 3 \dot{\phi} - 3 \h 
\left(\frac{\delta p}{\delta \rho} - w\right) \frac{\delta}{1 + w} \\ 
\label{eq:Euler}\nonumber
\dot{\theta} &=& -\h (1-3w)\theta-\frac{\dot{w}}{1 + w}\, \theta + 
\frac{\delta p}{\delta \rho}\, \vec{k}^2 \frac{\delta}{1 + w} + 
k^2 \psi \,, 
\end{eqnarray} 
where $\vec{k}$ is the wavevector, dots are derivatives 
with respect to conformal time, $\h=\dot a/a$ is the conformal Hubble 
parameter, $\delta \equiv \delta\rho/\rho$ is the density perturbation, 
$(\rho + p) \,\theta \equiv \imag k^j\delta T^0_{\,\,j}$ is the velocity 
perturbation, and $w = p/\rho$ is the equation of state.  These equations 
hold for each individual component, i.e.\ matter or dark energy.  

We define the effective (or rest frame) sound speed $c_s$ through 
(see, e.g., \cite{Hu98}) 
\begin{equation}
\frac{\delta p}{\rho} = c_s^2 \,\delta + 
3 \mathcal{H} (1 + w) (c_s^2 - c_a^2) \frac{\theta}{k^2}\,,
\end{equation}
where the adiabatic sound speed squared is 
\begin{equation}
c_a^2 \equiv \frac{\dot{p}}{\dot{\rho}} =
 w - \frac{1}{3\h} \frac{\dot{w}}{1 + w} \,. 
\end{equation}
In terms of $c_s$, Eqs.~(\ref{eq:continuity}) and (\ref{eq:Euler}) read
\beqa 
\label{eq:deltacs}
\frac{\dot{\delta}}{1 + w} &=&
3 \mathcal{H} (w - c_s^2) \frac{\delta}{1 + w} \\ 
&\qquad&- \left[k^2 + 9 \mathcal{H}^2 (c_s^2 - c_a^2)\right] 
\frac{\theta}{k^2} + 3 \dot{\phi} \nonumber\\ 
\label{eq:thetacs}
\frac{\dot{\theta}}{k^2} &=& 
(3 c_s^2 - 1) \mathcal{H} \frac{\theta}{k^2} + 
c_s^2 \frac{\delta}{1 + w} + \psi \,. 
\eeqa 

One can readily see that the source term in a $\ddot\delta$ equation 
will have a negative term involving $c_s^2 k^2$ from $\dot\theta$ (take the 
derivative of Eq.~\ref{eq:deltacs} and substitute in Eq.~\ref{eq:thetacs}), 
indicating that growth is suppressed on small scales, $k>\h/c_s$.  However, 
perturbations will exist in the dark energy density even for $c_s=1$, 
albeit at a very low level within the Hubble scale $k>\h$.  As $c_s$ drops 
below unity, the suppression is itself suppressed and inhomogeneities 
in the dark energy can be sustained.  All such perturbations will vanish 
though as $1+w\to0$, regardless of $c_s^2$.
In the combination of Eqs.~(\ref{eq:deltacs})
and (\ref{eq:thetacs}) into a single second order equation
for $\delta$, the terms involving the metric in this equation
are all proportional to $1 + w$ (or derivatives thereof)
so that in the limit $1 + w \to 0$ the perturbations decouple from the
metric and do not experience a gravitational force leading to growth.

The dark energy perturbations affect the metric perturbations, and thus the
perturbations in the matter, through the Poisson equation 
\begin{equation}
\label{eq: poisson}
k^2 \phi = -4 \pi G a^2 \sum_i \rho_i \left(\delta_i + 
3 \h (1 + w_i) \frac{\theta_i}{k^2}\right),
\end{equation}
where the sum runs over all components.  
For a perfect fluid, there is no anisotropic stress so $\psi=\phi$. 

\begin{figure}
  \begin{center}{
  \includegraphics[width=\columnwidth]{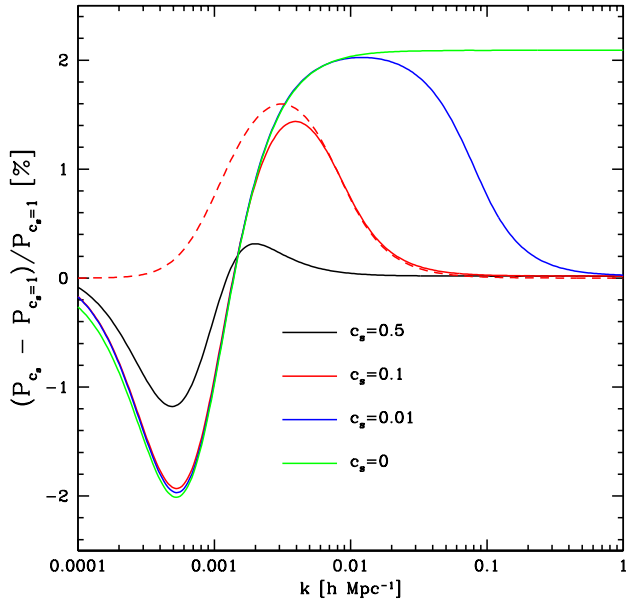}
  }
  \end{center}
  \caption{The deviation of the power spectrum of the matter density
    perturbations (Newtonian gauge) from the $c_s=1$ case is plotted vs.\
    wavenumber $k$.  Three regions -- above the Hubble scale (small $k$), 
below the sound horizon (large $k$), and the transition in between -- can 
clearly be seen.  The models have $w=-0.8$ (deviations will be smaller 
for $w$ closer to $-1$) and constant sound speed as 
labeled. For the $c_s=0.1$ case, we also show the result (dashed curve) in
terms of the gauge invariant variable $D_g$ as defined in
\cite{Durrer01} (in that work $\Phi$ is equal to minus our $\phi$).
This illustrates that the low $k$ behavior
is strongly gauge dependent.
}
  \label{fig:psdiffcs}
\end{figure}

Therefore we expect the density power spectrum to be affected by the 
dark energy sound speed in distinct ways on different scales.  On 
superhorizon scales, $k<\h$, the density power spectrum becomes 
independent of the dark energy sound speed.  Here the perturbations are 
determined by the curvature fluctuation \cite{Bardeen80, Bert06}. 
Between the Hubble scale and the sound horizon, $\h\ls k\ls\h/c_s$, 
a sound speed $c_s<1$ will enhance the density inhomogeneities (modulo 
gauge dependence around the Hubble scale).  Finally, 
on smaller scales, $k\gs\h/c_s$, inhomogeneity growth is always suppressed 
and the sound speed becomes irrelevant.  We illustrate these behaviors in 
Figure~\ref{fig:psdiffcs}.  (All power spectra in this paper are for linear 
theory and shown at $a=1$, and are calculated using
CAMB \cite{LewChalLas00} and CMBeasy \cite{DorMul04, Doran05}.)  Note that 
the strength of the deviation from the $c_s=1$ 
behavior is a steep function of $c_s$ for $c_s\approx0.1$.

\section{Dark Energy Models}
\label{sec: models}

We study three classes of dark energy models to elucidate the role of 
sound speed and $1+w$, from early to late times. 

{\bf 1) Constant $w$ models.} We begin with the simplest model of dark energy
with sound speed different from the speed of light: a constant equation of
state $w$ and a constant sound speed $c_s$.  This is mostly for historical
comparison to \cite{BeanDore04}, since the current constraint on constant
equation of state is $w=-0.97\pm0.08$ \cite{Union2} (using only geometric data
independent of the sound speed) and so the effects of sound speed are
suppressed due to $1+w\approx 0$.

\begin{figure}
  \begin{center}{
  \includegraphics*[width=\columnwidth]{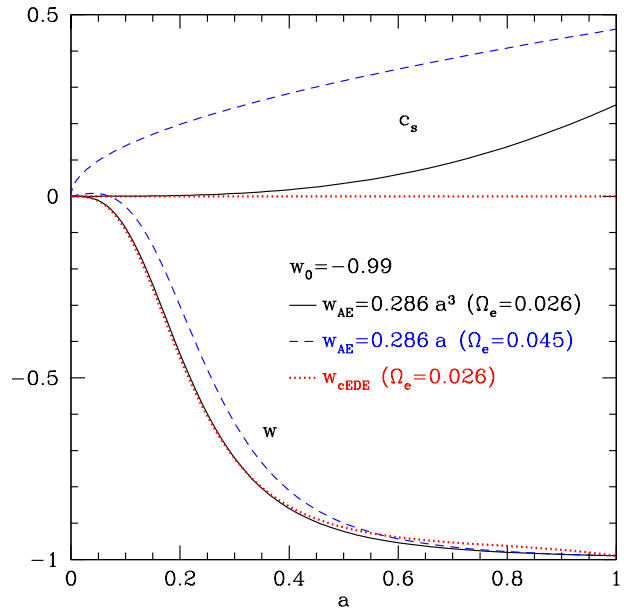}
  }
  \end{center}
  \caption{The equation of state (lower three curves) and sound speed 
(upper three curves) as a function of scale factor are illustrated for 
two models.  The aether model takes $s=3$ (solid curves) or $s=1$ (dashed 
curves) and $w_0=-0.99$; the early dark energy density $\Omega_e$ is 
determined from these parameters.  Note that the cEDE model (dotted curves, 
also taking $w_0=-0.99$, and here setting $c_s=0$) is a close match to 
the aether model. 
}
  \label{fig:cswa}
\end{figure}

{\bf 2) Early dark energy with constant speed of sound (cEDE).} In order to
allow for a period where $w$ is further from $-1$ and so the sound speed has 
more influence, we also consider a model with varying equation of state but
constant sound speed.  We choose the phenomenological early dark energy model
of \cite{DorRob06} but allow $c_s$ to be a free (constant) parameter.  At
early times $w$ approaches 0 in this model and so the value of $c_s$ can have
observational consequences.  The model parameters are the fraction of dark
energy density at early times $\Omega_e$ (this approaches a constant), the
equation of state today $w_0$, and $c_s$.  We call this generalization the
cEDE model.  Here
\begin{eqnarray}
\ode(a)&=&\frac{\ode-\ome\,(1-a^{-3w_0})}{\ode+\om a^{3w_0}} 
+\ome\,(1-a^{-3w_0}) \\[0.2cm] 
w(a)&=&-\frac{1}{3[1-\ode(a)]} \frac{d\ln\ode(a)}{d\ln a} 
\end{eqnarray} 
where the current dark energy density $\ode=1-\om$. In this model, 
$c_s={\rm const}$.  We show an example of $w(a)$ in Figure~\ref{fig:cswa}. 

{\bf 3) Barotropic (``aether'') dark energy models.} 
The third model we treat
  is a particular case of the barotropic class of dark energy, where there is
  an explicit relation determining the pressure as a function of energy
  density.  Several physical models for the origin of dark energy fall in this
  class, and have attractive properties as discussed below.  

Ref.~\cite{linsch} showed that all such viable models could be
  written as a sum of an asymptotic constant energy density $\rhoi$ (with
  $w_\infty=-1$) and a barotropic fluid, or aether, with positive equation of
  state $w_{AE}>0$.  The sound speed is completely determined by $w_{AE}$ and
  has the property that $c_s^2\le w_{AE}$.  Moreover, to admit an early matter
  dominated era, $w_{AE}(a\ll1)\to0$, and hence $c_s^2(a\ll1)\to0$.  We adopt 
the form  $w_{AE}=\beta a^s$ so
  \begin{eqnarray} 
    \rho_{de}(a)&=&\rho_\infty+\rho_{AE}(a)\\[0.2cm] 
    \rho_{AE}(a)&=&\rho_{AE,0}\,a^{-3}\,e^{3\beta(1-a^s)/s} \\[0.05cm]
    w(a)&=&-\frac{\rho_\infty}{\rhoi+\rho_{AE}(a)} 
    +w_{AE}(a)\,\frac{\rho_{AE}(a)}{\rhoi+\rho_{AE}(a)} \\[0.2cm] 
    c_s^2(a)&=&w_{AE}(a)-\frac{s}{3} \frac{w_{AE}(a)}{1+w_{AE}(a)} \,, 
  \end{eqnarray} 
  where $\rhoi=\rho_{de,0}-\rho_{AE,0}$.  There are two free parameters in
  addition to the dark energy density today: $\beta$ and $\rho_{AE,0}$ -- 
one less than in the cEDE case (we will fix $s=3$ usually).  
Note that the effective early dark energy density 
  $\ome\approx(\rho_{AE,0}/\rho_{m,0})\,e^{3\beta/s}$ and the present 
equation of 
  state is $w_0=-1+(\rho_{AE,0}/\rho_{de,0})\,(1+\beta)$.  As discussed by
  \cite{linsch}, the barotropic model strongly ameliorates the coincidence
  problem, motivating why $w\approx-1$ today.

Our three models thus span constant $w$ and constant $c_s$, varying 
$w$ and constant $c_s$, and varying $w$ and varying $c_s$ (but with 
$c_s$ determined by $w$).  We illustrate their equation of state and 
sound speed behaviors in Figure~\ref{fig:cswa}.  
We expect a cEDE early dark energy 
model with $c_s=0$ to show the greatest effect of sound speed on the 
observables.  Since cEDE can look so much like the barotropic model, 
in $w(a)$ and more approximately in $c_s$, we do not treat the barotropic 
model separately in the following sections, but rather consider it as a 
motivation for cEDE.  The barotropic model possesses the advantage of having 
$c_s=0$ at early times (and $w_0\approx-1$ at late times) being determined 
by physics rather than being adopted as phenomenology.

\section{Impact on Cosmological Observations \label{sec:cmbgal}} 

We now consider angular power spectra of cosmological observables that 
are sensitive to the speed of sound of dark energy, with the aim of 
comparing the predictions to current observations (so we do not here 
include higher order correlations, leaving for future work such 
signatures and their effect on constraining non-Gaussianity). 

\subsection{Angular Power Spectra}

The matter density fluctuations, potential fluctuations, and the radiation
field are influenced by the dark energy sound speed as discussed in
Sec.~\ref{sec:depert}.  From these we can form, and measure, the angular auto-
and cross-power spectra.  We consider the CMB temperature anisotropy power
spectrum, the galaxy (or other large scale structure tracer) overdensity 
field in a redshift bin $i$, and their
crosscorrelation, giving the power spectra $C_l^{XY}$, where
$\{XY\}=\{TT,Tg_i,g_i g_j\}$.  See the Appendix for a review of how the power
spectra relate to the potential power spectrum.

\begin{figure}
  \begin{center}{
  \includegraphics*[width=\columnwidth]{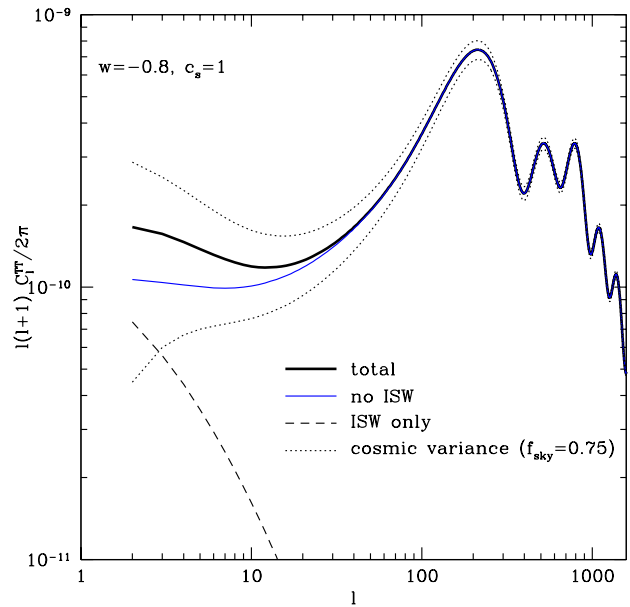}
  }
  \end{center}
  \caption{CMB temperature power spectrum for $w=-0.8$ and $c_s=1$, 
explicitly showing the contribution of the late-time ($z<10$) ISW effect.
}
  \label{fig: TT}
\end{figure}

\begin{figure*}
  \includegraphics[width=\columnwidth]{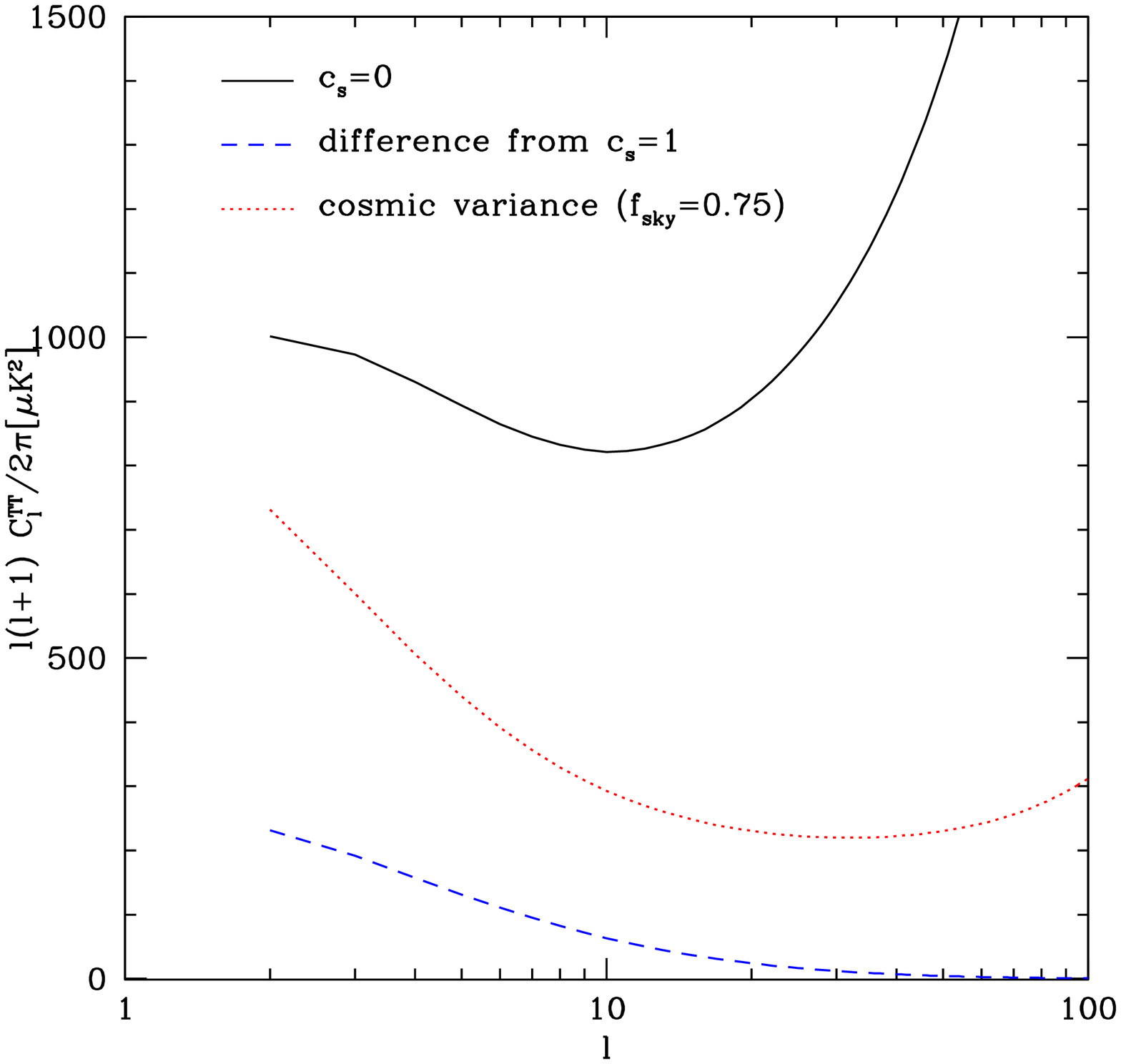}
  \includegraphics[width=\columnwidth]{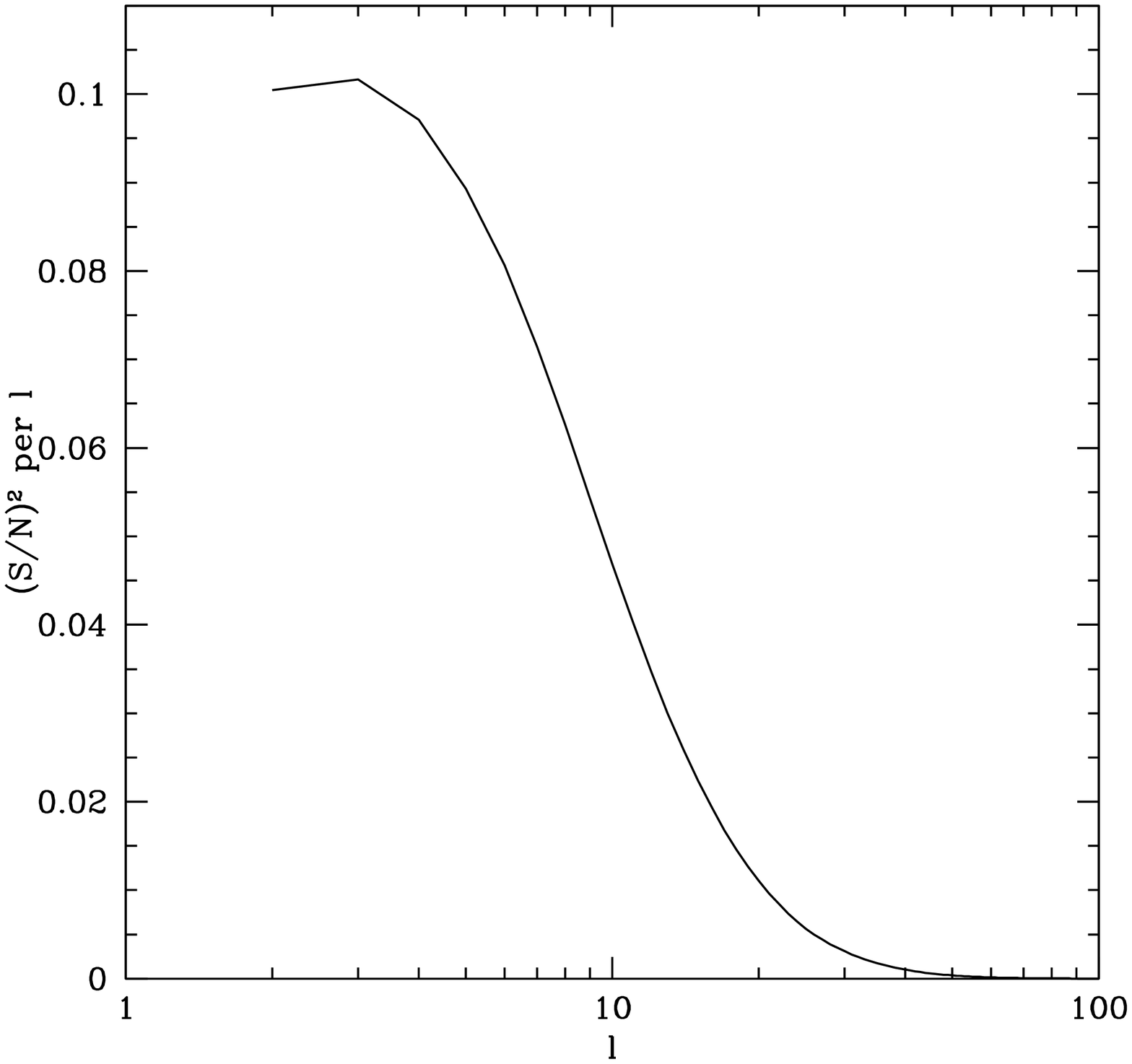}
  \caption{Left panel: CMB temperature power spectrum for $c_s=0$, and its
    difference from the $c_s=1$ case, are plotted for $w=-0.8$, along with the
    cosmic variance.  Right panel: The signal relative to the noise (here 
just cosmic variance) 
is low, with the total summed over all multipoles $S/N \simeq 1.0$. 
    Compensating the difference between the models by varying the other
    cosmological parameters would make the $S/N$ even smaller.  }
  \label{fig: TT S2N}
\end{figure*}

Figure~\ref{fig: TT} shows a typical temperature power spectrum. The signal 
from the sound speed dependence enters through the ISW effect, which is 
also plotted separately in the figure.  The extra power from the ISW effect 
arises from the decay of the potential as the dark energy impacts matter 
domination at late times; in the concordance model the cosmological 
constant dark energy causes a decay in the potentials of about $25\%$ 
between the matter dominated era and the present. While the decay arises 
from the change in the expansion history due to the dark energy equation 
of state, it can be ameliorated by increased dark energy clustering
if the dark energy sound speed is small.  Figure~\ref{fig: TT S2N} 
illustrates the influence of the sound speed.

The ISW effect can be measured 
\cite{boughn04,nolta04,fosalba03,scranton03,fosalba04,Padmanabhan:2004fy,afshordi04a,cabre06,Hoetal08,Giannantonio08} and one might hope to constrain 
the sound speed in this way.  However, since the effect occurs only on 
the largest angular scales, cosmic variance swamps the signal.  This is 
demonstrated in the left panel of Fig.~\ref{fig: TT S2N} for a cosmic variance 
limited experiment scanning 3/4 of the whole sky.  The right panel 
explicitly displays the low signal-to-noise for each multipole, with the 
difference between $c_s=0$ and $c_s=1$ only amounting to $S/N=1$ when 
summed over all multipoles.

\begin{figure}
  \begin{center}{
  \includegraphics*[width=\columnwidth]{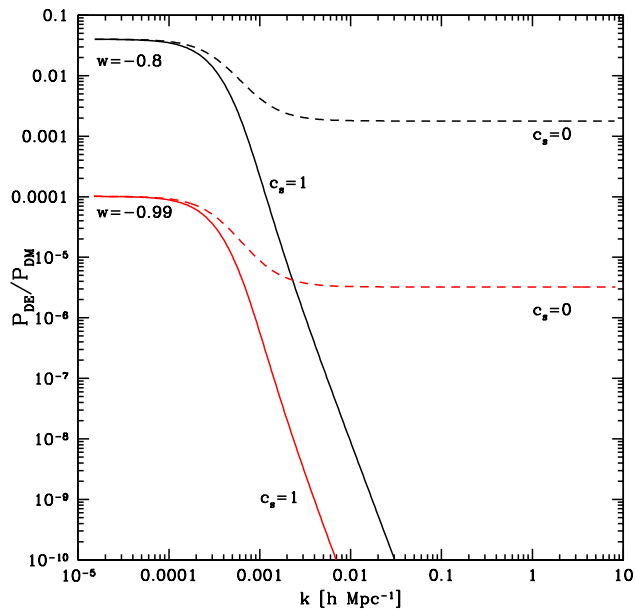}
  }
  \end{center}
  \caption{The ratio of the dark energy to dark matter density power spectra 
(Newtonian gauge) is plotted for various values of constant $w$ and $c_s$. 
Although $c_s=0$ gives dramatically more power on subhorizon scales than 
$c_s=1$, the direct ratio of the  dark energy power to the matter 
power is negligible. 
}
  \label{fig:cdmDEratio}
\end{figure}

For the galaxy or matter density fluctuations, the dark energy sound speed can
have a larger effect.  Note that the dark energy perturbations themselves
remain small relative to the matter inhomogeneities, despite a low sound speed
having a dramatic effect on the dark energy clustering.
Figure~\ref{fig:cdmDEratio} shows that on superhorizon scales the level of
dark energy power is $(1 + w)^2$ relative to the dark matter power (because at
superhorizon scales the perturbations remain adiabatic and the ratio
$\delta_{DE}/\delta_{DM} = 1 + w$).  On subhorizon scales, the ratio depends
strongly on the dark energy sound speed.  For $c_s=0$, the ratio is
scale independent in the subhorizon regime: during matter domination, one can
show analytically that then 
\begin{equation}
\frac{P_{DE}}{P_{DM}} = \left (\frac{1 + w}{1 - 3 w}\right)^2 \quad
({\rm matter\,\, dominated})
\end{equation}
but this ratio becomes smaller by roughly a factor of two by today.  For a
canonical sound speed $c_s=1$, the dark energy power is strongly suppressed
relative to the dark matter power, with the ratio scaling as $k^{-4}$.

\begin{figure*}
\includegraphics[width=\columnwidth]{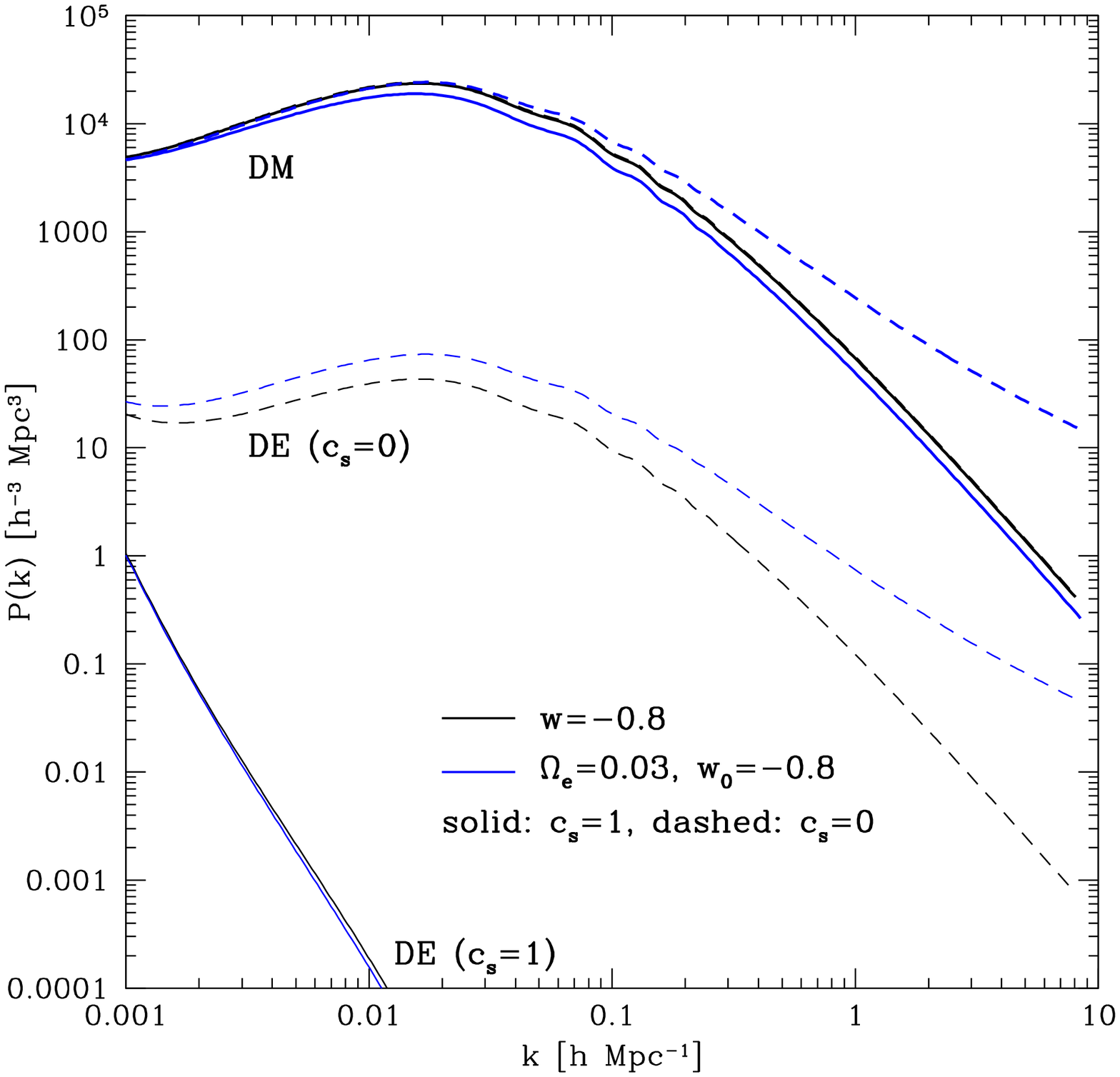}
\includegraphics[width=\columnwidth]{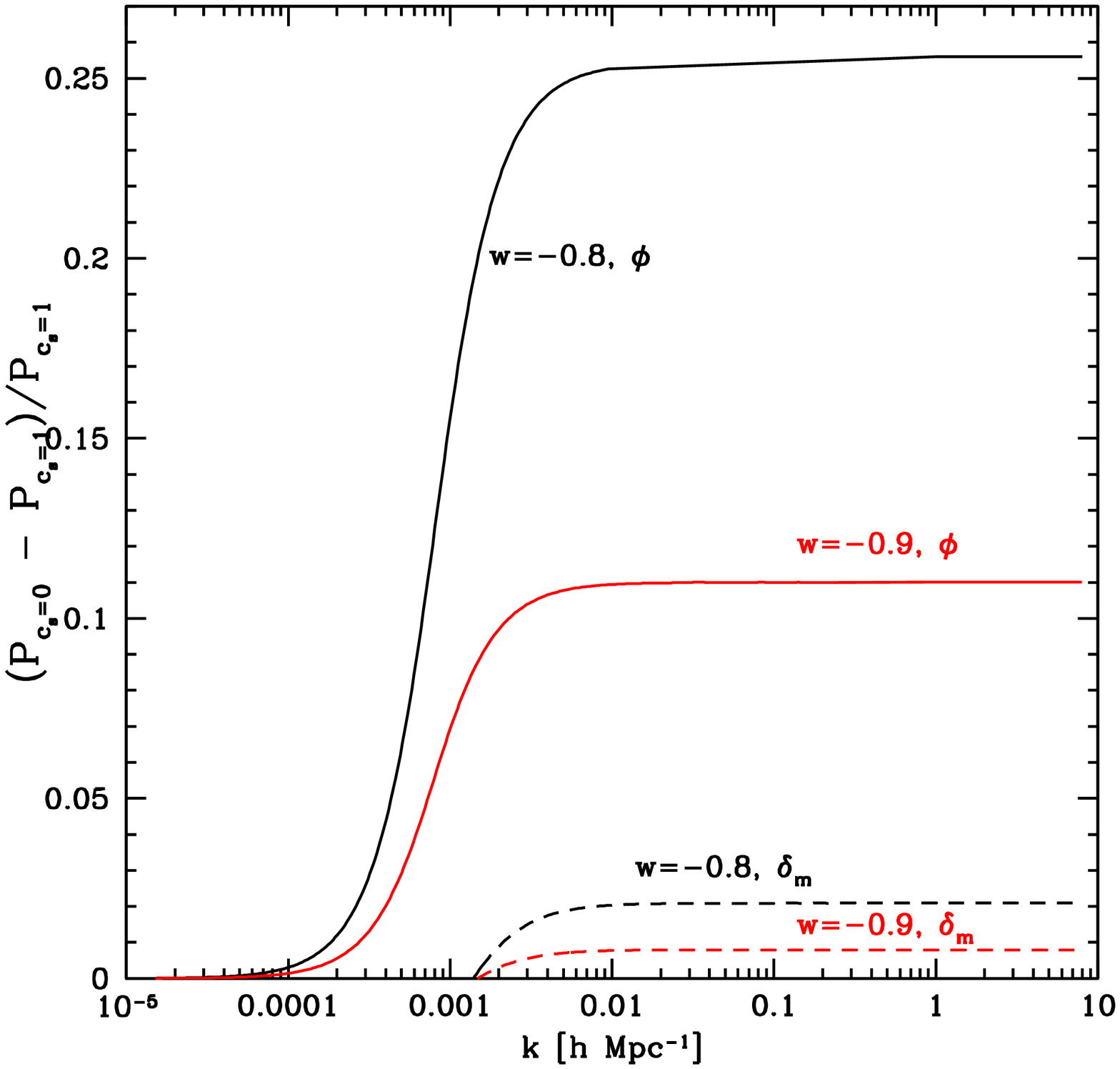}
  \caption{Left panel: dark energy (lower four, thin curves) 
    and dark matter (upper, thick curves) density power spectra
    for different choices of the dark energy equation of state and sound
    speed. Right panel: relative differences in the potential ($\phi$) 
and matter density ($\delta_m$) power
    spectra between $c_s=0$ and $c_s=1$ (matter and dark energy perturbations
    in Newtonian gauge).  }
  \label{fig: PS}
\end{figure*}

The matter power spectrum itself, however, is affected by the dark energy
sound speed through the potential perturbations induced by the dark energy
inhomogeneities.  Figure~\ref{fig: PS} shows in the left panel the 
{\it absolute\/} comparison of the dark matter and dark energy power (in 
contrast to 
the relative difference between the two in Fig.~\ref{fig:cdmDEratio}).  On
this log scale one cannot see the influence of the dark energy sound speed on
the dark matter power, so the right panel plots the deviation with respect to
the $c_s=1$ case.  We see that the deviation due to $c_s=0$ is at the percent
level in the matter density power and the tens of percent level in the
potential perturbation power.

The density and potential are related through the Poisson equation.  For
example, for $w=-0.8$ and $c_s=0$, the amplitude of the dark energy
perturbations is about $4 \%$ of the dark matter perturbation (i.e.\ the power
ratio is about $1.6\times10^{-3}$ on subhorizon scales as seen from
Fig.~\ref{fig:cdmDEratio}).  According to the Poisson equation, Eq.~(\ref{eq:
  poisson}), this translates into about a $12 \%$ increase in $\phi$ going
from $c_s=1$ to $c_s=0$, because today $\rho_{\rm DE} \approx 3 \rho_m$ and
because in the $c_s=1$ case the dark energy contribution to the Poisson
equation is negligible.  Hence, as shown in the right panel of
Fig~\ref{fig: PS}, we get about a $25\%$ increase in the power spectrum of
$\phi$. 

Note that the (late) ISW effect is proportional to the {\it change} in
potential $\Delta \phi$ between matter domination and today. In the standard
case, this decay is about $1/4$ of the potential during matter domination and
thus about $1/3$ of the potential today, i.e.~$\Delta \phi \equiv \phi_0 -
\phi_{MD} \approx - \frac{1}{4} \phi_{MD} \approx -\frac{1}{3} \phi_0$.
Hence, the change in the potential at present of $12 \%$ due to enhanced 
dark energy clustering corresponds to a change in the ISW effect of 
approximately $3 \times 12 \% = 36 \%$ (i.e.\ 
in $[\Delta\phi(c_s=0)-\Delta\phi(c_s=1)]/\Delta\phi(c_s=1))$. 
This enhancement gives the ISW effect extra sensitivity to dark energy
clustering relative to other probes.

The matter density perturbation is of course also affected, but with only 
about a $1\%$ increase in its amplitude.  This effect on the potential 
today through the Poisson equation is therefore subdominant to the direct 
effect of the dark energy perturbation itself.

Now that we have seen the basic effects of the dark energy sound speed and
equation of state on the observables, we consider the specific instances of
the constant $w$ model and cEDE model.  We can already guess that to obtain 
reasonable constraint on the sound speed we will want a model that has as
large a $1+w$ and as small a $c_s$ as is consistent with the observations, for
a substantial part of cosmic history.

\begin{figure*}[!t]
 \includegraphics[width=\columnwidth]{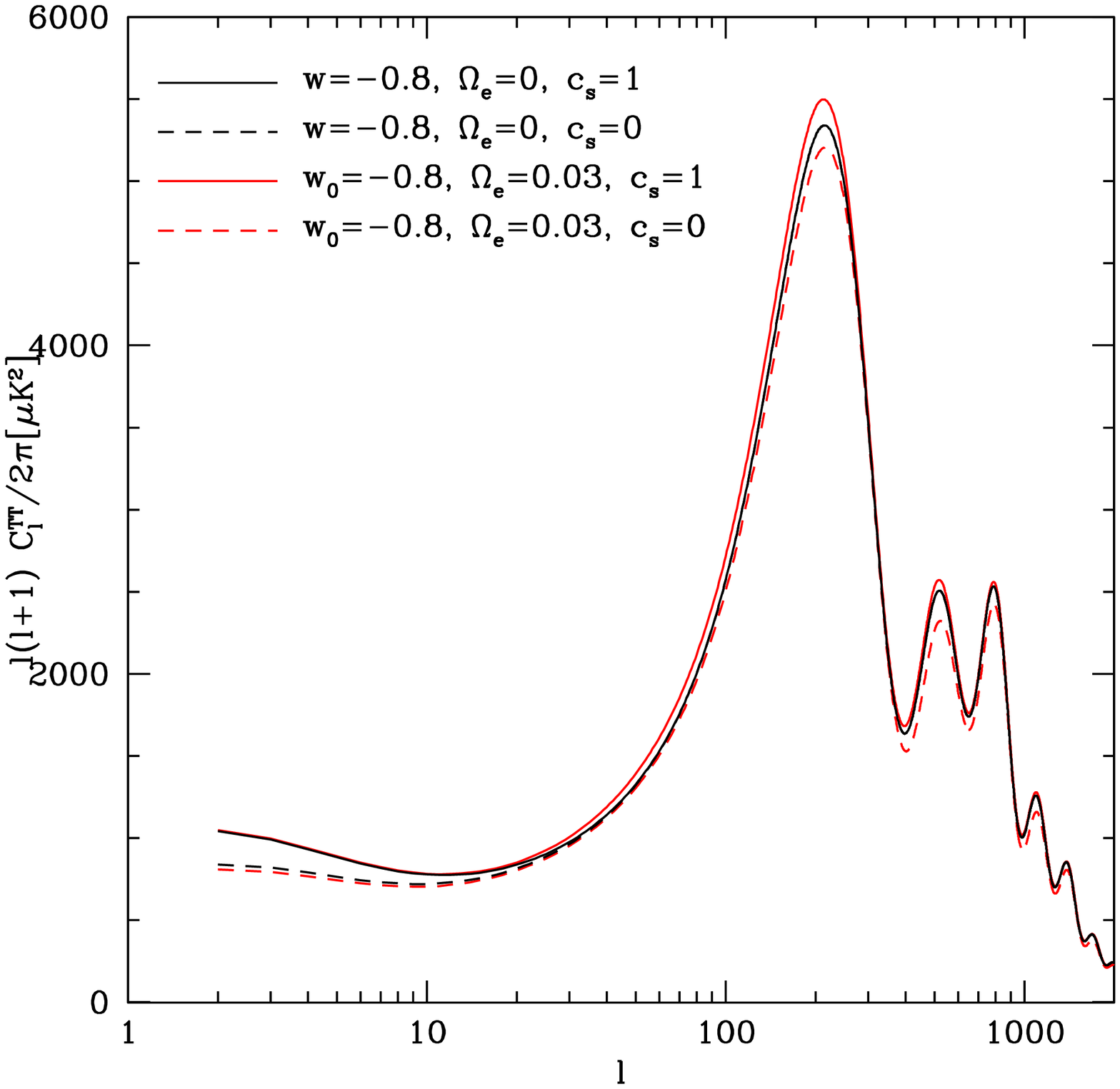}
 \includegraphics[width=\columnwidth]{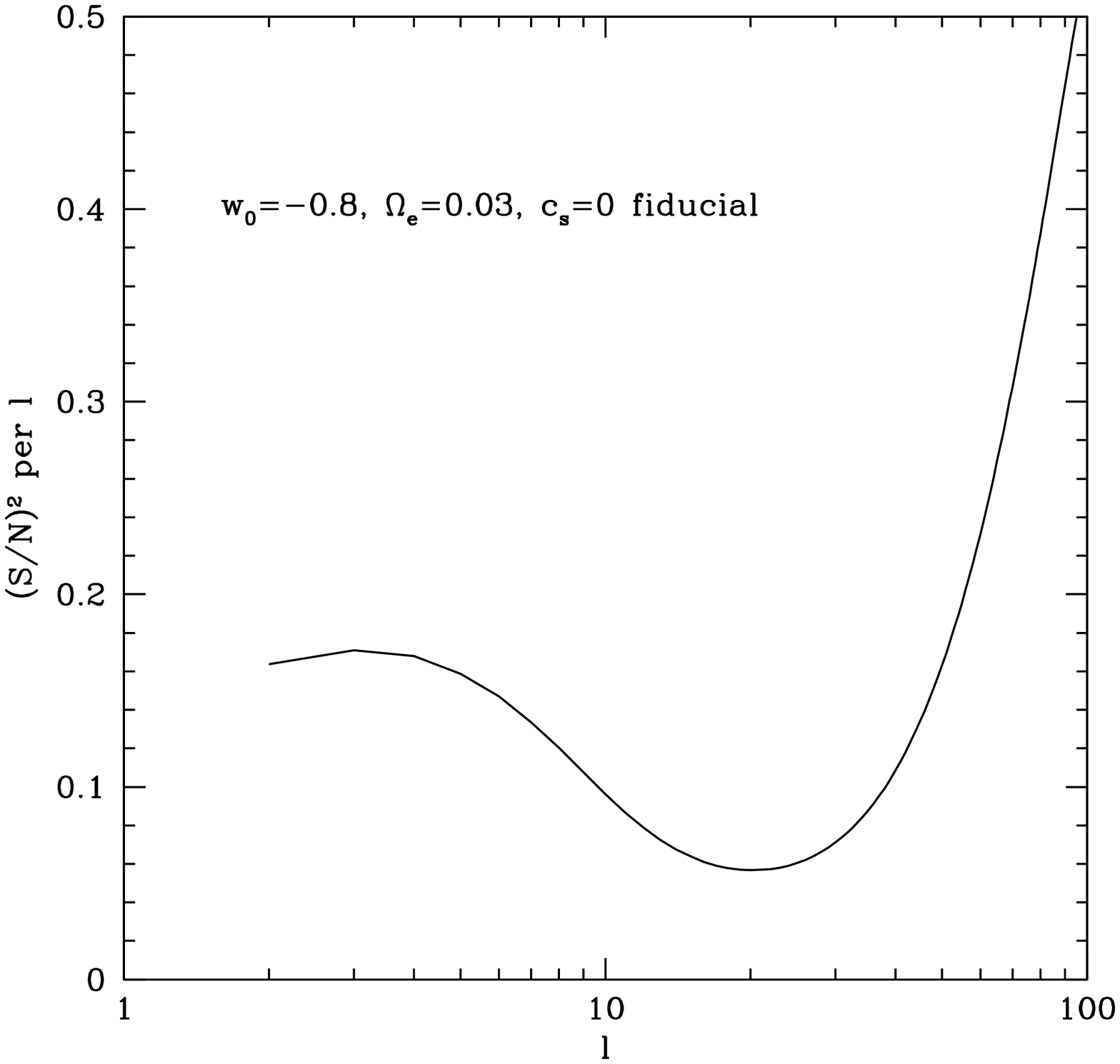}
 \caption{Left panel: CMB temperature spectra for the early dark energy cEDE
   model with $\Omega_e=0.03$, $w_0=-0.8$ is plotted for $c_s=0$ and $1$.  The
   effect of changing the sound speed on the late ISW effect is a little
   stronger than in the case of ordinary $w=-0.8$ dark energy (also shown),
   but the major difference comes from higher $\ell$, where the early dark
   energy exhibits significant differences between $c_s=0$ and $c_s=1$, while
   ordinary dark energy does not.  Right panel: Signal to noise squared per 
mode 
   for distinguishing $c_s=1$ from the $c_s=0$ fiducial is plotted vs.\
   multipole.  The late ISW (treated as $\ell<21$) contributes only $\sn2 =
   1.8$; including higher $\ell$, say all $\ell \leq 2000$, gives $\sn2 =
   8.8\times 10^3$. However, the differences at high $\ell$ can at least
   partly be compensated by varying other cosmological parameters.  
}
  \label{fig: TT ede}
\end{figure*}

\subsection{Estimating constraints in constant $w$ Model \label{sec:constw}}

We begin by estimating the chances of constraining the sound speed using the 
$\chi^2$ between two extremes: $c_s=0$ and $c_s=1$. Since we consider angular
power spectra and crosscorrelations of observables on the sky (labeled by 
capital letters below), $\chi^2$ is in general given by 
\begin{equation}
\chi^2 = \sum_\ell \sum_{\{XY\}, \{ZW\}} \Delta C_\ell^{XY}\,
({\bf Cov}_\ell)^{-1}_{XY, ZW} \,\Delta C_\ell^{ZW},
\label{eq:chisq}
\end{equation}
where $\Delta C_\ell^{XY}$ is the difference in spectra between the two
cases and the covariance is given by
\begin{equation}
({\bf Cov}_\ell)_{XY, ZW} = \frac{1}{(2\ell + 1) f_{\rm sky}} 
\left (\tilde{C}_\ell^{XZ}\tilde{C}_\ell^{YW}+\tilde{C}_\ell^{XW} \tilde{C}_\ell^{YZ} \right )\,,
\end{equation}
with
\begin{equation}
\tilde{C}^{XY}_\ell = C^{XY}_\ell + N^{XY}_\ell\,,
\end{equation}
where $f_{\rm sky}$ is the fraction of the sky that is observed, $C_\ell^{XY}$
are the fiducial spectra and $N_\ell^{XY}$ are the noise power spectra so that
$\tilde{C}_\ell$ are the observed power spectra that include the noise.
(See the Appendix for further details.)  
For the $\chi^2$ estimates of this section we only consider the CMB 
temperature power spectrum and we will consider the cosmic variance 
dominated limit where the noise power spectrum is much smaller than the 
fiducial power spectrum, $N_\ell^{TT} = 0$. Hence, Eq.~(\ref{eq:chisq}) 
simplifies to
\beq
\chi^2 = \ha f_{\rm sky} \sum_\ell (2 \ell + 1)  \left(\frac{\Delta C_\ell^{TT}}{C_\ell^{TT}}\right)^2.
\eeq

Assuming Gaussian likelihood, the quantity $\chi^2$ is equivalent to the 
signal to 
noise squared with which we can distinguish $c_s=1$ from our fiducial $c_s=0$
if all the other parameters were known exactly. Since in reality we should
marginalize over the other parameters as well, $\chi^2$ is an {\it upper\/}
bound on the signal to noise squared for distinguishing the two sound speeds.
Therefore if we find a low $\chi^2$ then there is little hope of constraining
$c_s$ with the assumed dataset.  To amplify the chances of detection, we 
examine 
$w=-0.8$, since in the limit $w \to -1$ dark energy perturbations become
irrelevant regardless of the value of the value of the sound speed; given that
$w=-0.8$ is already an unlikely value given current data, the calculated
signal to noise squared $(S/N)^2$ could be an overoptimistic estimate of the
true value.

Figure~\ref{fig: TT S2N} confirms that the discrimination between sound speeds
through the CMB temperature autocorrelation is poor, as discussed in the
previous subsection.  Cosmic variance swamps the difference between even the
extremes, $c_s=0$ and $c_s=1$, and the total $(S/N)^2\approx1$.  Note that 
this took 
cosmic variance to be calculated from the most optimistic case, $c_s=0$,
where the noise is significantly lower, so one truly cannot determine $c_s$
with the CMB temperature anisotropy despite all the most optimistic
assumptions.

The overall significance of the mere existence of the ISW effect (i.e.\ the
$\chi^2$ between the CMB power with the ISW effect artificially removed and
the true CMB) is only $\sn2_{ISW}=3.7$.  The potential decay in a model with
dark energy sound speed $c_s = 0$ is a little less than half the contribution
in the $c_s=1$ case, thus explaining the $\sn2_{\Delta c_s=1}\approx
(1/4)\sn2_{ISW}=1.0$ quoted above. Thus the ISW signal in the CMB temperature
spectrum is too blunt a tool to explore dark energy sound speed.

We must go beyond the CMB temperature spectrum to consider the galaxy-galaxy
power and temperature-galaxy crosscorrelation data. 
Rather than proceeding further with halfway measures such as calculating the
signal to noise to determine whether we would be able to place to constraints
on $c_s$ while fixing all other parameters, we instead carry out a full
likelihood analysis in Sec.~\ref{sec:data}.

\subsection{Estimating constraints in cEDE model}

In the early dark energy case, we find that the ISW signal in both the
CMB temperature autocorrelation and temperature-galaxy crosscorrelation 
is comparable to the signal in the case of ordinary dark energy (which 
typically has an energy density fraction relative to matter of $\sim10^{-9}$ 
at CMB last scattering).  However, there is another source of distinction. 
Dark energy in the cEDE model has $w\approx0$ at CMB last scattering; if 
in addition $c_s=0$, then cEDE behaves at early times just like dark matter, 
with significant clustering of the dark energy. 
This will affect not only the large scale, late time ISW contribution to 
the CMB but also the early Sachs-Wolfe effect and the acoustic peaks. 

Therefore we expect a clearer observational signature of the sound speed
than for ordinary dark energy.  Figure~\ref{fig: TT ede} shows the effect
of changing the sound speed in the cEDE model.  The CMB temperature 
autocorrelation alone delivers $\sn2\approx 9 \times 10^3$ (for 
$\ell_{\rm max}=2000$).  This seems more promising for constraining the 
sound speed, and again we proceed to a full likelihood analysis.

\section{Measuring the Speed of Darkness \label{sec:data}}

To obtain accurate constraints on the dark energy sound speed we perform a
Markov Chain Monte Carlo (MCMC) likelihood analysis over the set of parameters
$\{\log c_s, p_{\rm dark}, \omega_b, \omega_c, \Omega_{de}, \tau, A_s, n_s
\}$, where $p_{\rm dark}$ is either $w$, in the constant $w$ case, or
$\{w_0,\Omega_e \}$, in the cEDE case, $\omega_b = \Omega_b h^2$ is the
present physical baryonic energy density density, $\omega_c = \Omega_c h^2$ 
is the present physical cold dark matter energy density, $\Omega_{de}$ is 
the present relative
energy density in the dark energy, $\tau$ is the reionization optical depth,
$A_s$ the amplitude of primordial scalar perturbations (defined relative to a
pivot scale of $k=0.05\,{\rm Mpc}^{-1}$) and $n_s$ is the spectral index of the
primordial scalar perturbations.  Note that we choose $\log c_s$ as the sound
speed parameter because most of the sensitivity is at small values of $c_s$.

For current data we include the CMB temperature power spectrum from WMAP5
\cite{Komatsu09}, the crosscorrelation of these temperature anisotropies with
mass density tracers including the 2MASS (2-Micron All Sky Survey), SDSS LRG
(Sloan Digital Sky Survey Luminous Red Galaxies), SDSS quasars, and NVSS (NRAO
VLA All Sky Survey) radio sources, following \cite{Hoetal08}, and the SDSS LRG
autocorrelation function from \cite{Tegetal2006}.  To break degeneracies with
background cosmology parameters and constrain the expansion history, we use
the supernovae magnitude-redshift data from the Union2 compilation
\cite{Union2}.

\begin{figure}
  \begin{center}{
  \includegraphics*[width=\columnwidth]{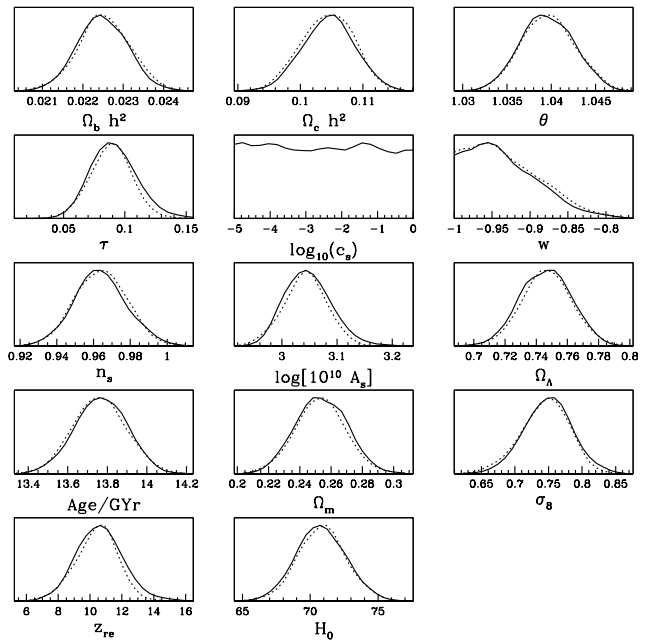}
  }
  \end{center}
  \caption{Constant equation of state case, plotting the 
    marginalized one dimensional probability distributions using
    data from supernovae (Union2), CMB (WMAP5), galaxy autocorrelation (SDSS LRG), and the cross correlation between large scale structure
    tracers (see text) and CMB temperature anisotropies. Solid lines are for
    the model with $\log(c_s)$ a free parameter (with a flat prior), whereas
    the dotted lines correspond to fixed $c_s=1$. 
 }
  \label{fig: w current}
\end{figure}

\begin{figure}
  \begin{center}{
  \includegraphics*[width=\columnwidth]{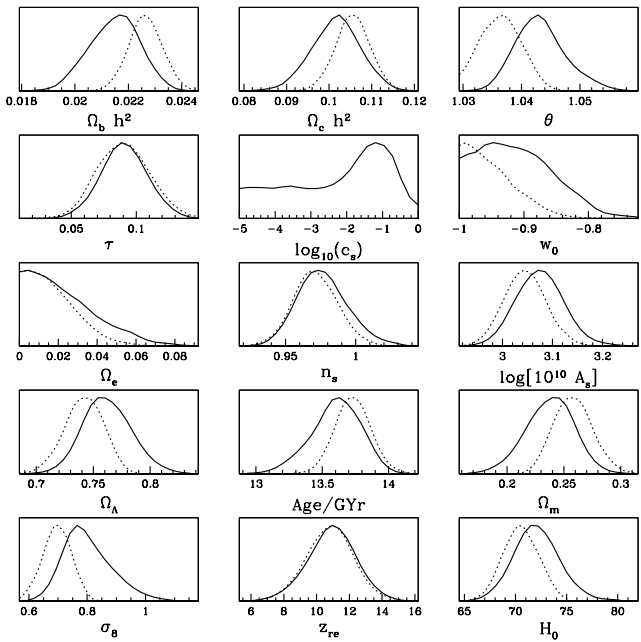}
  }
  \end{center}
  \caption{Early dark energy case, plotting the marginalized one
    dimensional probability distributions using data from supernovae (Union2),
    CMB (WMAP5), galaxy autocorrelation (SDSS LRG), and the cross
    correlation between large scale structure tracers (see text) and CMB
    temperature anisotropies. Solid lines are for the model with $\log(c_s)$ a
    free parameter (with a flat prior), whereas the dotted lines correspond to
    fixed $c_s=1$. 
}
  \label{fig: ede current}
\end{figure}

The MCMC package COSMOMC \cite{LewisBridle02} is used to calculate the joint
and marginalized likelihoods.  The results for the marginalized 1D probability
distributions are shown in Fig.~\ref{fig: w current} for the constant equation
of state case and in Fig.~\ref{fig: ede current} for the early dark energy,
cEDE case.  Dotted lines show the distributions when one fixes $c_s=1$.

In the constant $w$ case, no constraint can be placed on the sound 
speed, as expected from our earlier arguments.  In addition, the other 
parameter distributions are essentially unaffected by the value of $c_s$. 
For the cEDE case, however, some preference appears for a low sound speed, 
$c_s\lesssim0.1$, and this propagates through to the other parameters. 
Since early dark energy with a low sound speed acts like additional 
dark matter at early times, this allows a lower true matter density.  

It is intriguing to consider whether the apparent preference of current data
for the $\Lambda$CDM model is merely a consequence of overly restricting the
degrees of freedom of dark energy, and that instead a dark energy with
dynamics ($w_0\approx-0.95$), microphysics ($c_s\approx 0.04$), and long-time
presence ($\Omega_e\approx 0.02$) could be the correct model. 

\begin{figure}
  \begin{center}{
  \includegraphics*[width=8.6cm]{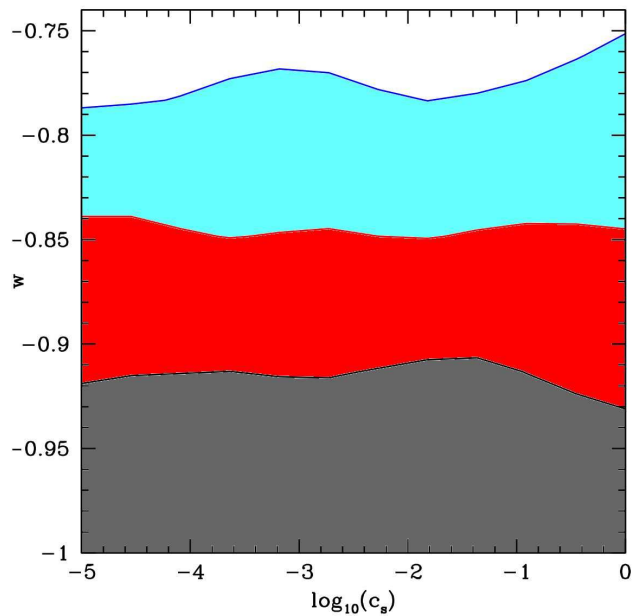}
  }
  \end{center}
  \caption{68.3, 95.4 and 99.7\% confidence level contours in the dark energy
    model with constant equation of state.  The constraints are based on
    current data including CMB, supernovae, LRG power
    spectrum and crosscorrelation of CMB with matter tracers.}
  \label{fig: cont constw}
\end{figure}

\begin{figure*}
  \includegraphics[width=5.8cm]{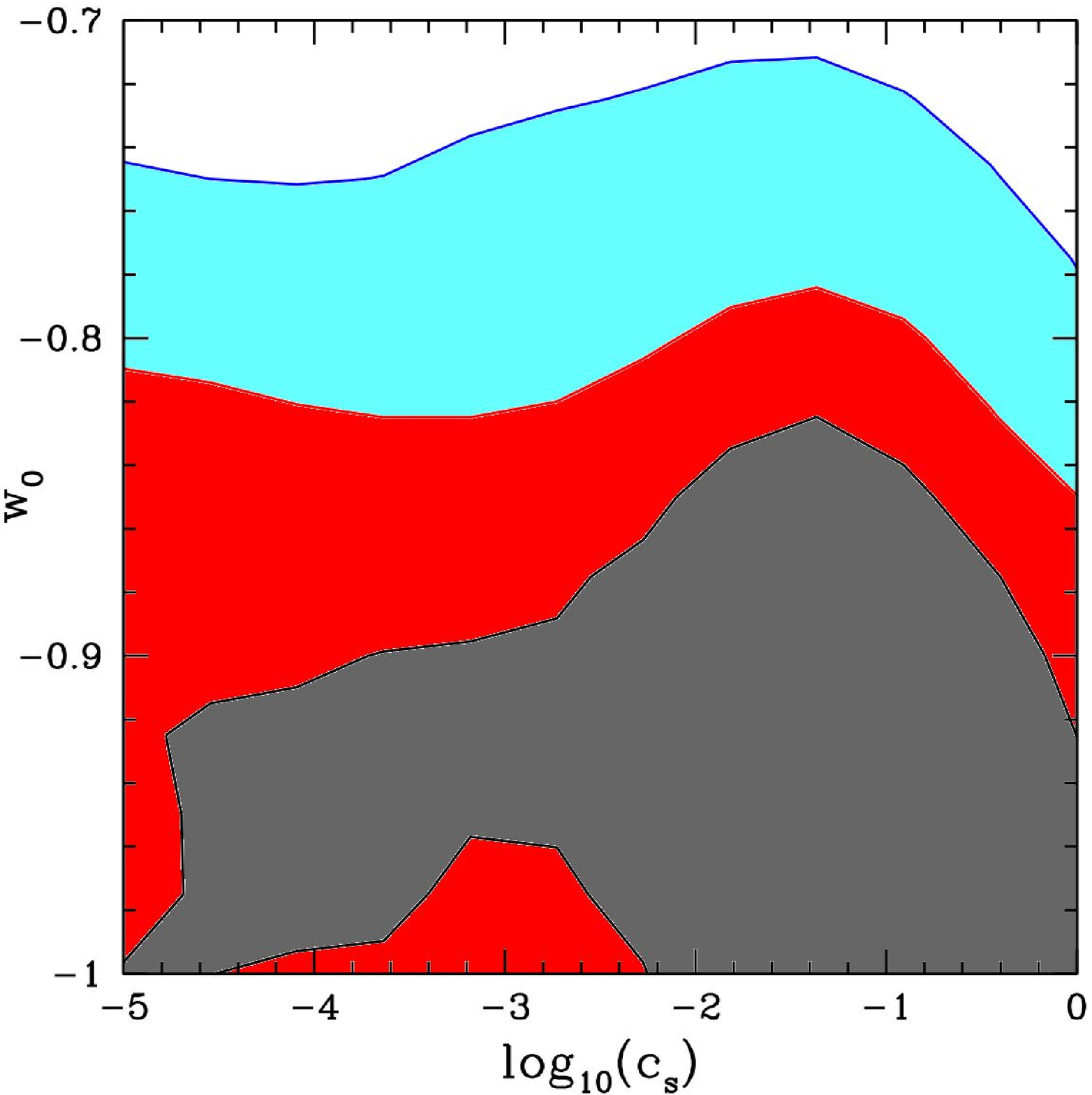}
  \includegraphics[width=5.8cm]{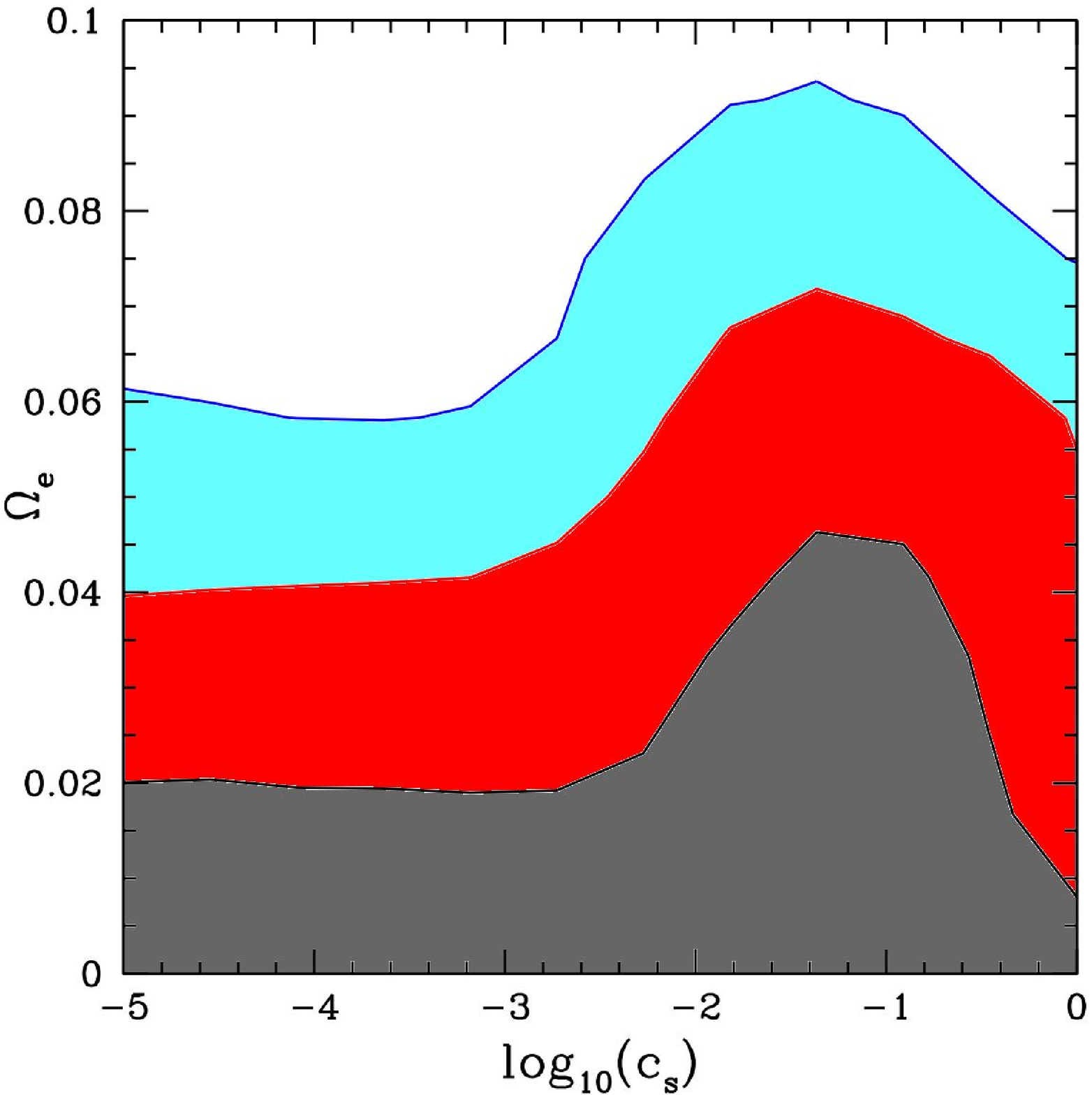}
  \includegraphics[width=5.8cm]{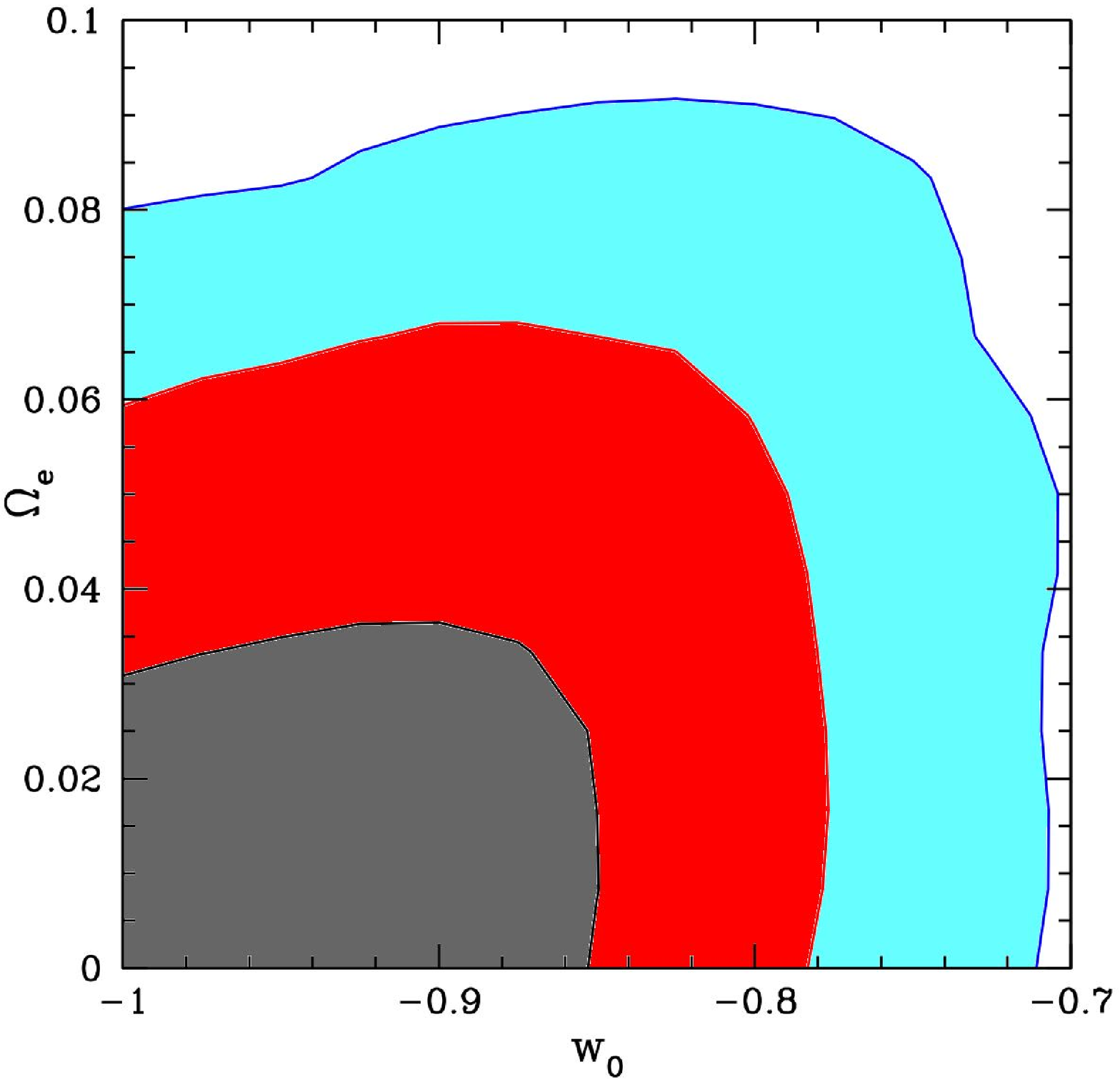}
  \caption{68.3\%, 95.4\% and 99.7\% confidence level contours in the 
cEDE early 
    dark energy model in the $w_0$-$\log c_s$ (left), $\ome$-$\log c_s$
    (middle) and $\ome$-$w_0$ (right) planes. The constraints are based on
    current data including CMB, supernovae, LRG power spectrum, and 
crosscorrelation 
    of CMB with matter tracers.}
  \label{fig: cont ede}
\end{figure*}

Figure~\ref{fig: cont constw} shows the 68.3\%, 95.4\% and 99.7\% confidence
level contours in the $w$-$\log c_s$ plane for the constant $w$ model.
We see 
that current data in this model prefer $w\approx-1$ but are completely
agnostic regarding $c_s$.  For the cEDE model, Fig.~\ref{fig: cont ede} shows
the joint probability contours among the dark energy parameters, in the
$w_0$-$\log c_s$, $\ome$-$\log c_s$, and $\Omega_e$-$w_0$ planes, with all
other parameters marginalized.  Here we see that the
model mentioned above, $(w_0,c_s,\Omega_e)=(-0.95, 0.04,0.02)$, is completely
consistent with the data, as is the cosmological constant $(-1,1,0)$.  It 
will be interesting to see how 
the best fit evolves with future data.

\section{Conclusions} 

Current cosmological data are in excellent agreement with the standard 
$\Lambda$CDM universe, with equation of state $w=-1$.  Nevertheless, 
the current data are also consistent with a wide variety of richer 
physics.  It is not clear that it is wise to assume that the physical 
explanation for dark energy in the universe is indeed given by restriction 
to a 
spatially smooth, constant in time energy density: the cosmological constant.  
Even after allowing for dynamical dark energy, there could 
be further degrees of freedom -- ``hidden variables'' or microphysics -- 
in the dark energy sector, harbingers of deeper physics that have not 
yet shown clear signatures in the data. An explicit search for these 
signatures, and thus the physics behind dark energy, should be near the 
top of the list of current efforts in cosmology.

In this paper we search for degrees of freedom beyond quintessence by 
examining the influence of the sound speed of dark energy, and its 
resulting spatial clustering of dark energy, on key observables and in 
current data.  
This extends earlier analyses, quantifying the effects on the dark 
matter and dark energy density perturbation power spectra, the potential 
power spectrum, and their crosscorrelation.  Where possible, we give simple 
scalings with $1+w$ and $c_s$.  We also explore models with time 
varying equation of state and sound speed. 

In the standard model with negligible dark energy at high redshift, 
the speed of sound is essentially not distinguishable with current data (see 
Fig.~\ref{fig: w current}) because current data favor $w\simeq -1$, and the
effects of clustering of dark energy vanish in this limit.  As $w$ gets 
further from $-1$, the influence of the sound speed increases; for models 
with $w\approx0$ at high redshift there is also the possibility of 
non-negligible amounts of early dark energy density.  Even just a couple 
percent of the total energy density in early dark energy dramatically 
improves the prospects for detecting dark energy clustering.  One can view 
the early dark energy fraction $\Omega_e$ as another degree of freedom 
to explore.  Indeed, carrying out a MCMC analysis we find in 
Figs.~\ref{fig: ede current} and \ref{fig: cont ede} that a model with 
dynamics, microphysics, and persistence: 
$(w_0,c_s,\Omega_e)=(-0.95, 0.04,0.02)$ is completely consistent with the
current data 
(although $\Lambda$ remains consistent as well). 

Discovery of the accelerating universe 12 years ago has propelled the 
physical interpretation of dark energy into one of the most important, 
exciting, and difficult problems in physics.  Although current 
observations indicate that the equation of state, as a constant or 
broadly averaged over time, is close to $-1$, this leaves considerable 
room for further physics, as demonstrated here using recent data.  To go 
further we should explore all three frontiers of the dynamics $w(a)$, 
the microphysics $c_s$ and spatial inhomogeneities, and the persistence 
$\Omega_e$.

\acknowledgments 

We are extremely grateful to the Supernova Cosmology Project for permission 
to use the Union2 supernova data before publication and for use of 
their computer cluster. RdP thanks Jeff Anderson for 
invaluable computer support and Marina Cort{\^e}s for useful conversations
about MCMC. EL and RdP have been
supported in part by the Director, Office of Science, Office of High Energy
Physics, of the U.S.\ Department of Energy under Contract
No.\ DE-AC02-05CH11231, and EL by the World Class University grant
R32-2008-000-10130-0.  DH is supported by the DOE OJI grant under contract
DE-FG02-95ER40899, NSF under contract AST-0807564, and NASA under contract
NNX09AC89G.

\appendix 

\section{Angular Power Spectra: Definitions} \label{appB}

Here we review how the observable power spectra of various quantities on the
sky are related to the three-dimensional primordial power spectrum and the 
transfer functions. 
We consider the CMB temperature anisotropies and galaxy
overdensities in redshift slices, or populations, labeled with the 
subscript $j$, and 
write the observables in direction $\hat{n}$ as line of sight projections 
along comoving radial coordinate $\chi$,
\begin{equation}
\label{eq: source}
X(\hat{n}) = \int d\chi \, S^X(\hat{n} \chi, \tau_0 - \chi),
\end{equation}
with $S^X(\vec{x}, \tau)$ the ``source term'' as a function of comoving 
position and conformal time ($\tau_0$ is the age
of the universe in conformal time).
Here $X$ represents the observable, which
could be a galaxy overdensity $g_j$ in the $j$th 
redshift bin or a CMB temperature anisotropy $T$.  For the galaxy
overdensity $g_j$, the source is
\begin{equation}
S^{g_j}(\vec{x}, \tau) = H(z(\tau)) \frac{n_j(z(\tau))}{n_j^A} b_j \delta_m(\vec{x}, \tau),
\end{equation}
where $n_j(z)dz$ is the average angular galaxy density of galaxy population
$j$ in the redshift interval $(z, z + dz)$, $n_j^A=\int dz\,
n_j(z)$ 
is the total average 
angular galaxy density of population $j$, and $b_j$ is the galaxy bias 
relative to the matter overdensity of bin $j$.
The Hubble factor $H(z)$ arises because the source was defined 
in terms of an integral 
over $\chi$ while $n_j(z)/n_j^A$ is normalized to unity in terms of an 
integral over $z$.

For CMB temperature anisotropies, the (Fourier transform
of the) source is given in Eq.~(12) of \cite{SelZal96}. 
The Integrated Sachs-Wolfe contribution to the CMB anisotropy is
nonzero when the universe is {\it not} matter dominated, and thus the
gravitational potentials $\phi$ and $\psi$ are not constant. The ISW source is
given by
\begin{equation}
\label{eq: ISW cource}
S^{\rm ISW}(\vec{x}, \tau) = \dot{\phi}(\vec{x}, \tau) + \dot{\psi}(\vec{x}, \tau),
\end{equation}
where dots denote derivatives with respect to conformal time.

If we expand the anisotropy field in spherical harmonics, $X(\hat{n}) =
\sum_{\ell m} a^X_{\ell m} Y_{\ell m}(\hat{n})$, the expansion coefficients
are given by
\begin{eqnarray}
\label{eq:alm1}
a^X_{\ell m} &=& \int d\Omega \, Y_{\ell m}^*(\hat{n}) X(\hat{n}) \nonumber \\
&=& (2\pi)^{-3/2}\int d\Omega \, Y_{\ell m}^*(\hat{n}) \int d^3 \vec{k} 
 \int d\chi \, e^{i \vec{k} \hat{n} \chi} S^X(\vec{k}, \tau_0 - \chi) \nonumber \\
&=& \sqrt{\frac{2}{\pi}}\, i^\ell \int d^3\vec{k} \, Y_{\ell m}^*(\hat{k}) 
\int d\chi \, j_\ell(k \chi) S^X(\vec{k}, \tau_0 - \chi),
\end{eqnarray}
where we have Fourier expanded
\begin{equation}
\label{eq: Fourier conv}
S^X(\vec{x}, \tau) = \int \frac{d^3 \vec{k}}{(2\pi)^{3/2}} \, 
e^{i\vec{k} \vec{x}} S^X(\vec{k}, \tau)\,,
\end{equation}
and we have used the Rayleigh plane-wave expansion
\begin{equation}
e^{i \vec{k} \cdot \hat{n} \chi} = 4\pi\sum_{\ell,m}i^\ell
j_\ell(k\chi)Y_{\ell m}^*(\hat{k})Y_{\ell m}(\hat{n})\,,
\end{equation}
where the $j_\ell$ is the spherical Bessel function. 
We now write $S^X(\vec{k}, \tau) = \psi_i(\vec{k}) \, S^X(k, \tau)$ where
$\psi_i(\vec{k})$ is the initial potential perturbation and $S^X(k, \tau)$ is
the source for $\psi_i = 1$, i.e.~it is a transfer function. Due to the
assumption of homogeneity, the transfer function does not depend on the
direction of the wavenumber, but only on its magnitude $k\equiv |\vec{k}|$. 
The statistics of the initial perturbations are given by
\begin{equation}
\left< \psi_i(\vec{k}) \psi_i(\vec{k}')\right> = 
P^{\psi}_i(k) \,\delta^{(3)}(\vec{k} + \vec{k}'),
\end{equation}
where $P^{\psi}_i(k)$ is the primordial potential power spectrum.  Assuming
statistical isotropy, the angular correlations between two quantities on the 
sky $X$ and $Y$ (where they may or may not be the same)  is given by the 
angular power spectrum
\begin{equation}
\label{eq: Clal}
\left<a_{\ell m}^X a_{\ell' m'}^{Y*}\right> = C^{XY}_\ell \delta_{\ell \ell'} \delta_{mm'} 
\end{equation}
where, using Eq.~(\ref{eq:alm1}), 
\begin{eqnarray}
\label{eq: Cl}
C^{XY}_\ell &=& \frac{2}{(2\pi)^2} \, \int d^3 \vec{k}\,
P^{\psi}_i(k)  \int d\chi \, j_\ell(k\chi) S^X(k, \tau_0 - \chi)\times \nonumber \\
&&\int d\chi' \, j_\ell(k \chi') S^Y(k, \tau_0 - \chi')\,.
\end{eqnarray}
In this work, we are specifically interested in the combinations $\{XY\} =
\{TT, T g_i, g_i g_j\}$, but Eq.~(\ref{eq: Cl}) is the general expression
for angular power or crosscorrelation spectra.

When the sources $S^X$ and $S^Y$ vary slowly compared to the spherical Bessel
functions in Eq.~(\ref{eq: Cl}), the triple integral can to a good
approximation be reduced to a single integral. Setting 
$P(k)=P(k=(\ell+1/2)/\chi(z))$ and using the 
asymptotic (for $\ell\gg 1$) formula
that $\left(2/\pi\right)\int k^2 dk j_\ell(k\chi) j_\ell(k
\chi')=\left(1/\chi^2\right)\delta(\chi-\chi')$, we find
\begin{eqnarray}
C_\ell^{XY} &=& \frac{2 \pi^2}{\left(\ell+1/2\right)^3} \int d\chi \, \chi \,
\Delta^{\psi}_i\left(\frac{\ell +1/2}{\chi}\right) \times \nonumber \\[0.1cm]
&&S^X\left(\frac{\ell +1/2}{\chi}, \tau_0 - \chi\right)
\,S^Y\left(\frac{\ell +1/2}{\chi}, \tau_0 - \chi\right)
\end{eqnarray}
where $\Delta(k) \equiv k^3 P(k)/(2\pi^2)$.  We use the power spectra to
calculate the $\chi^2$ (signal-to-noise) in Eq.~(\ref{eq:chisq}).

Finally, we need to specify formulae for noise in the observed spectra 
$C_\ell^{XY}$.  The covariances between the spectra are given by
\begin{equation}
{\bf Cov}(C_\ell^{XY}, C_{\ell'}^{ZW}) = \delta_{\ell \ell'} \frac{1}{(2\ell + 1) f_{\rm sky}} 
\left (\tilde{C}_\ell^{XZ}\tilde{C}_\ell^{YW}+\tilde{C}_\ell^{XW} \tilde{C}_\ell^{YZ}
\right )\,,
\end{equation}
where
\begin{equation}
\tilde{C}^{XY}_\ell = C^{XY}_\ell + N^{XY}_\ell\,.
\end{equation}
Here $f_{\rm sky}$ is the sky coverage, $C_\ell^{XY}$
are the fiducial spectra and $N_\ell^{XY}$ are the noise power spectra.
For the galaxy density fields, the white noise power spectra are given by
\beq
N_\ell^{g_j g_j} = \frac{1}{n_j^A}\,,
\eeq
and for the CMB it is given by
\beq
N_\ell^{TT} = \Delta_T^2\, e^{\ell(\ell+1) \theta^2_{\rm FWHM}/(8\ln 2)}\,,
\eeq
where $\Delta_T$ is the sensitivity and $\theta_{\rm FWHM}$ is the full width
half max angle of the Gaussian beam. The noise cross power spectra can be 
assumed to vanish.
 
The treatment of the covariances for actual data is typically more 
complicated than the above.  In this paper, we use the covariances and 
treatment of the observables as given by the data packages in COSMOMC, 
\cite{Hoetal08, Tegetal2006, Komatsu09}
for the angular spectra, and the Union2 supernovae covariance matrix 
including systematics.

\bibliography{refs}

\end{document}